\documentstyle[preprint,pra,aps]{revtex}

\begin{document}
\draft

\title{Persistent Currents in the Heisenberg chain with a weak link}

\author{T.M.R. Byrnes, R.J. Bursill, H.-P. Eckle, C.J. Hamer}  

\address{School of Physics,    \\                                           
The University of New South Wales,        \\                            
Sydney, NSW 2052, Australia.}

\author{A. W. Sandvik}
\address{Department of Physics, \\
{\AA}bo Akademi University, \\
Porthansgatan 3, FIN-20500 Turku, \\
Finland}

\date{\today}

\maketitle 

\tightenlines

\begin{abstract}
	The Heisenberg chain with a weak link is studied, as a simple
example of a quantum ring with a constriction or defect. The Heisenberg
chain is equivalent to a spinless electron gas under a Jordan-Wigner
transformation. Using density matrix renormalization group and quantum
Monte Carlo methods we calculate the spin/charge stiffness of the model,
which determines the strength of the `persistent currents'. The
stiffness is found to scale to zero in the weak link case, in agreement
with renormalization group arguments of Eggert and Affleck, and Kane and
Fisher.
\end{abstract}


\section{INTRODUCTION} 

Technological advances in recent years have allowed the fabrication of
electrical and even mechanical devices on the nanometer scale,
where individual atoms or electrons can be manipulated.
The physics of these devices poses a plethora of fundamental questions 
through a rich variety of novel quantum effects\cite{moriond}. 
This has led to an upsurge of theoretical interest in 
the physics of `quantum wires',
`quantum dots' and more general physics at the mesoscopic
or nanometer scale\cite{simons}. 
The effects of 
electron-electron interactions are typically enhanced in systems of reduced 
dimensionality, leading to nonperturbative effects, such as the breakdown of
Fermi liquid behavior in one-dimensional metals, and single-electron
charging effects in quasi-zero-dimensional systems (quantum dots).
An important milestone in the nascent field of nanomechanics was the 
experimental discovery
that not only the electrical but also the {\em mechanical} properties of
metallic structures on the nanometer scale 
exhibit apparently universal nonmonotonic
quantum corrections \cite{exp1,exp2}, which could be explained theoretically
within the framework of a Jellium model \cite{stafford}.  

In the electrical domain, paradigm systems to investigate mesoscopic
behavior have long been small ring-shaped or multiply
connected devices, where the application of a magnetic flux piercing
the device leads to persistent currents. However, the theoretical
prediction of these persistent currents, first in superconducting and then
in normal conducting materials, has always predated experimental 
investigation, which has only become feasible in the last decade
\cite{efetov}. Very recently, with the discovery of a tunable Kondo
effect in quantum dots, the persistent currents of multiply connected
systems with magnetic quantum dots - Kondo rings for short - have
aroused considerable interest\cite{EJS,kang,SA}.

In this paper we explore a very simple system, which may serve to model
a metallic quantum wire ring with a weak junction, or {\em
constriction}. 
It consists of the standard spin-1/2 Heisenberg antiferromagnetic
spin chain,
which by a Jordan-Wigner transformation is equivalent 
to a spinless electron gas in one dimension, where
the exchange coupling is weakened at a single link. We use Density
Matrix Renormalization Group (DMRG) \cite{white92,gehring97} and Quantum Monte
Carlo (QMC) methods\cite{sandvik1} to obtain numerical results on chains of up
to 256 sites, and perform finite-size scaling extrapolations to the bulk
limit. We study the spin stiffness, which under the Jordan-Wigner
transformation is equivalent to the charge stiffness of the electron
gas and is related to the persistent current,
as a function of the weak link
coupling. We also study the spin correlations across the
weak link. 

Eggert and Affleck\cite{eggert} have previously
studied the Heisenberg chain with an isolated impurity using exact
diagonalization and conformal field theory techniques. They find that in
renormalization group language a single weak link across the ends of an
open chain corresponds to an irrelevant operator, and therefore the open
chain is a stable fixed point under such a perturbation. Thus they
predict that in the bulk limit a chain with a weak link will behave like
an open chain.
These findings for a concrete model system are in agreement with the general
predictions of Kane and Fisher\cite{kane} for the general one--dimensional
interacting electron gas, i.e. the Luttinger liquid.
An integrable version of the Heisenberg spin chain with defects
has also been studied\cite{schmitteckert,EPR,sydney}.
The case of a single defect corresponds to a weak link. 
However, as opposed to the case discussed here where only one bond is
modified, integrability requires a modification of two adjacent bonds
and an additional three--spin coupling. Although the defect of this
integrable chain is completely transparent to particle scattering,
the persistent current is renormalized by the defect strength.

Let us briefly review\cite{byers} how the persistent current arises, for the
simple case of free electrons. Start from the real-space continuum
Hamiltonian
\begin{equation}
H = -\frac{\hbar^{2}}{2m_{e}}\sum_{\alpha}\int_{0}^{L}dx \psi^{\dagger}_{\alpha}(x)
\partial_{x}^{2}\psi_{\alpha}(x)
\end{equation}
where $\psi_{\alpha}(x)$ is an electron field with spin index $\alpha =
\pm 1$, and $L$ is the circumference of the ring. Now thread the
ring with a magnetic flux $\Phi$, producing an Aharonov-Bohm
effect\cite{aharonov}. The quantum phase 
\begin{equation}
\frac{e}{\hbar c}\int_{0}^{L}A_{\mu}d_{\mu}x = \frac{e}{\hbar c}\Phi
\end{equation}
can be encoded via a gauge transformation in the twisted boundary
conditions
\begin{equation}
\psi_{\alpha}(L) = e^{i\phi}\psi_{\alpha}(0),
\end{equation}
where
\begin{equation}
\phi = 2\pi \frac{\Phi}{\Phi_{0}}
\end{equation}
and $\Phi_{0} = hc/e$ is the elementary flux quantum. 

When an Aharonov-Bohm field is applied, the Hamiltonian acquires the
usual interaction term (we set $\hbar = c = 1$ henceforth) 
\begin{equation}
H_{int} = - \int dx A_{\mu}J^{\mu}(x).
\end{equation}
Thus for a constant field $A_{1} = \Phi/L$ the corresponding
``persistent current" is given by the Feynman-Hellman theorem:
\begin{equation}
J^{1} \equiv I(\Phi) = -\frac{\partial
E_{0}}{\partial\Phi}
\end{equation}
which can be expanded
\begin{equation}
I(\Phi) = -D_{c}\frac{\Phi}{L} + O\big((\frac{\Phi}{L})^{2}\big)
\end{equation}
where $D_{c}$ is the ``charge stiffness". If we {\em assume} that
$I(\Phi)$ is purely linear in $\Phi$ (as can be proved for the pure
Heisenberg chain\cite{hamer}), then the charge stiffness and hence the
persistent current can be estimated from the difference in energy
between the system with anti-periodic boundary conditions ( $E_{0}^{-} =
E_{0}(\Phi = \Phi_{0}/2)$) and periodic boundary conditions ($(E_{0}^{+} =
E_{0}(\Phi = 0)$) 
\begin{equation}
\label{chargestiffness}
D_{c} = \frac{8L}{\Phi_{0}^{2}}[E_{0}^{-} - E_{0}^{+}]
\end{equation}
The corresponding quantity in the Heisenberg chain is the spin
stiffness.

In section II of the paper we briefly summarize the DMRG and QMC methods used 
to calculate this quantity.
In section III we present our results, and in section
IV our conclusions are summarized.

\section{Method}

We study the spin-1/2 Heisenberg quantum spin chain with a single weak 
coupling $ J' < J $ between two adjacent spins located between sites $ i = N $ and $ i = 1 $. The Hamiltonian is
\begin{equation}
H = J \sum_{i=1}^{N-1} {\bf S}_i \cdot {\bf S}_{i+1} + J' {\bf S}_N \cdot {\bf S}_1.
\end{equation}
There are a total of $ N $ sites in the ring. We study $ J'/J $ in the range $ [0,1] $, hence we have either open or periodic boundary conditions (OBC or PBC) at the extremities of this range. We also consider anti-periodic boundary conditions in the same range of $ J'/J $, to obtain the charge stiffness $ D_c $ according to (\ref{chargestiffness}). 
The model in these limits is exactly solvable by Bethe ansatz \cite{bethe}, and
 hence is often used as a testing ground for various DMRG methods 
\cite{white92,bursill99}. The quantities that have been calculated using DMRG 
include the ground state energy, the singlet and triplet gaps, and correlation 
functions. 

\begin{figure}
\centerline{\epsffile{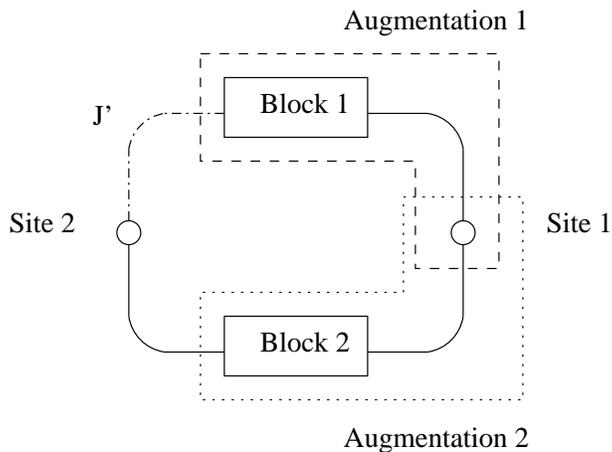}}
\caption{The augmentation process within one DMRG iteration. Augmentation 1 (Augmentation 2) gives the new Block 1 (Block 2) in the next DMRG iteration. }
\label{fig2}
\end{figure}

The ``infinite-lattice'' DMRG method \cite{white92} is used here, applied with 
periodic boundary conditions. The lattice is split into two blocks and two sites as shown in Fig. \ref{fig2}. The weak link is placed between block 1 and site 2. At any time the superblock consists of a system block and an environment block, plus two extra sites. 
The presence of the weak link destroys the translational invariance normally exploited in usual DMRG schemes, hence we cannot simply make a copy of the density matrix in one block and transfer it to the other block. Therefore in a single DMRG iteration, two density matrices are constructed (one for each block), and the basis set for each block originates from its corresponding density matrix. Each block increases in size by a single site in a single DMRG iteration. We calculate results for lattice sizes $ N = 4 $ to 64, in steps of two. The quantities calculated here are the ground state energies, and the correlation of the spins across the weak link. 
The total number of density matrix eigenstates retained in a block was $ m = 350 $ in the basis truncation procedure.

We also carried out quantum Monte Carlo simulations using the stochastic
series expansion (SSE) method\cite{sandvik1}. In this case, the spin
stiffness can be directly calculated as the second derivative of the
energy with respect to the phase, which is given in terms of the
fluctuation of the winding number in the simulations \cite{sandvik2}.

\section{Results}


To estimate the accuracy of the DMRG, we perform convergence tests with $m$, 
the number of basis states retained in a block. Table \ref{table1} shows sample
 results for $ J'/J = 0.5 $ and $ N = 64 $ with both periodic and anti-periodic
 boundary conditions. We see good convergence with the number of basis states 
retained: in particular, for the ground state energy we have convergence to 1 
part in $ 10^6 $. We obtained 
independent estimates using SSE techniques for the periodic 
case, which yields $ E_0/J = -28.2178(3)$, agreeing perfectly with the DMRG 
results. The correlations between the spins across the links have a marginally 
lower accuracy because of round-off error, but even here we 
have an error in the region of 1 part in $ 10^4 $. 

Since the SSE method is a finite-temperature quantum Monte Carlo
method we have to run the simulations at sufficiently low temperature
to converge the quantities of interest to their ground state values.
In Table \ref{table2} we show the convergence of the energy and the charge
stiffness for a 256-site chain. All SSE results discussed below were
obtained at inverse temperatures $\beta$ where the results do not
differ, within statistical errors, from results at $\beta/2$.

\begin{figure}
\centerline{\epsffile{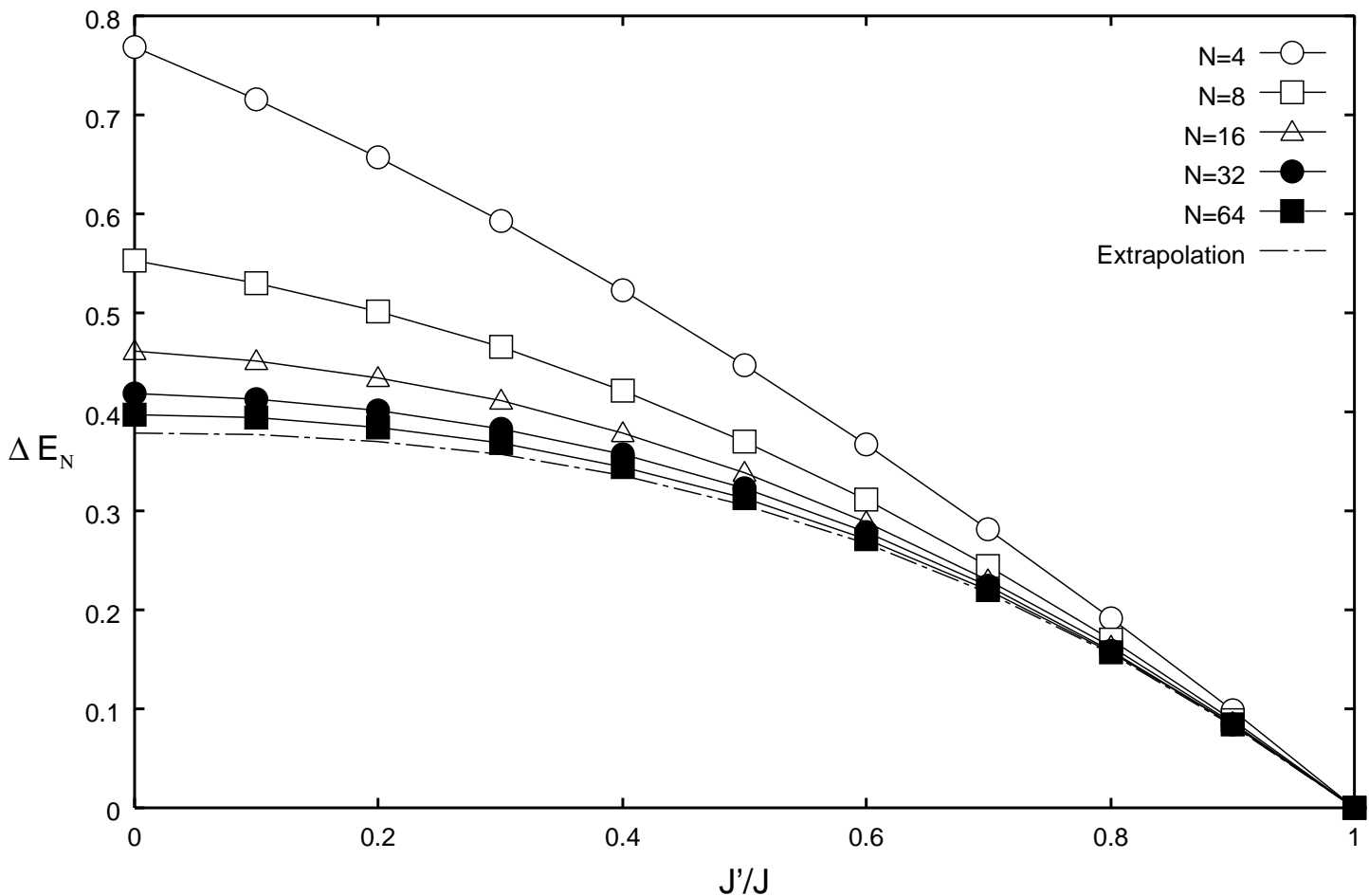}}
\caption{$ \Delta E_N = (E_0(N;J') - E_0(N;J'=J))/J $ as a function of $ J'/J $ for the ring with periodic boundary conditions. The data is extrapolated using a simple fit to obtain the bulk limit. }
\label{fig5}
\end{figure}

\begin{figure} 
\centerline{\epsffile{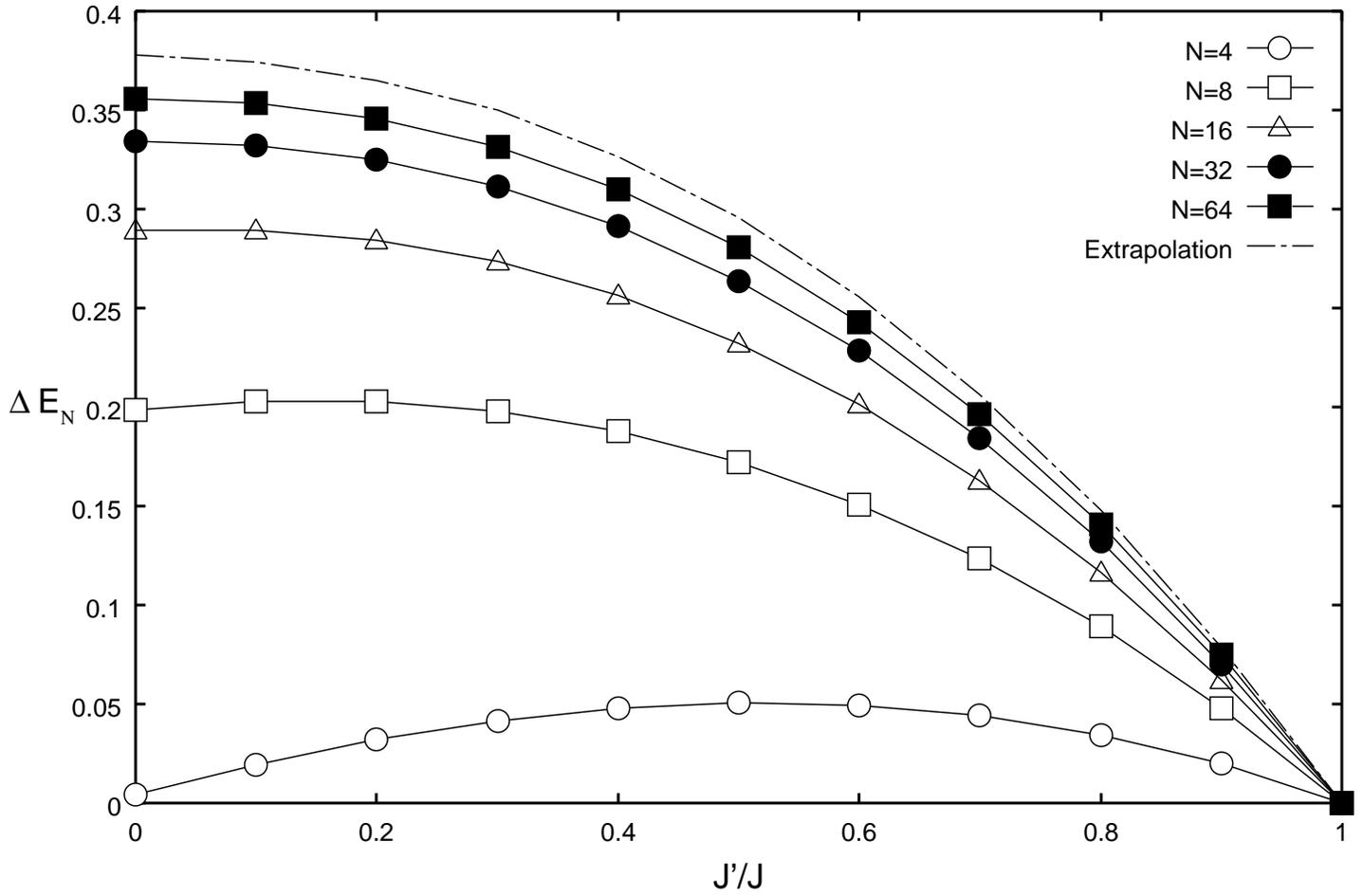}}
\caption{As for Fig. \ref{fig5}, but with anti-periodic boundary conditions. }
\label{fig6}
\end{figure}

Ground state energies were calculated for the periodic and anti-periodic 
rings, as a function of the weak link coupling $ J'/J $. 
Figs. \ref{fig5} and \ref{fig6} shows DMRG estimates for the quantity $\Delta E_N= (E_0(N;J') - E_0(N;J'=J))/J $ as a function of
$J'/J$, for several different lattice sizes up to $N = 64$. An
extrapolation in $ 1/N $ can be performed to extract the bulk limit for each 
value of $ J' $ by a simple polynomial fit to the data, giving us the 
extrapolated curve for the periodic and anti-periodic cases. It can be seen that for the periodic case the bulk values are approached from above, while for the anti-periodic case the limit is approached from below.

\begin{figure} 
\centerline{\epsffile{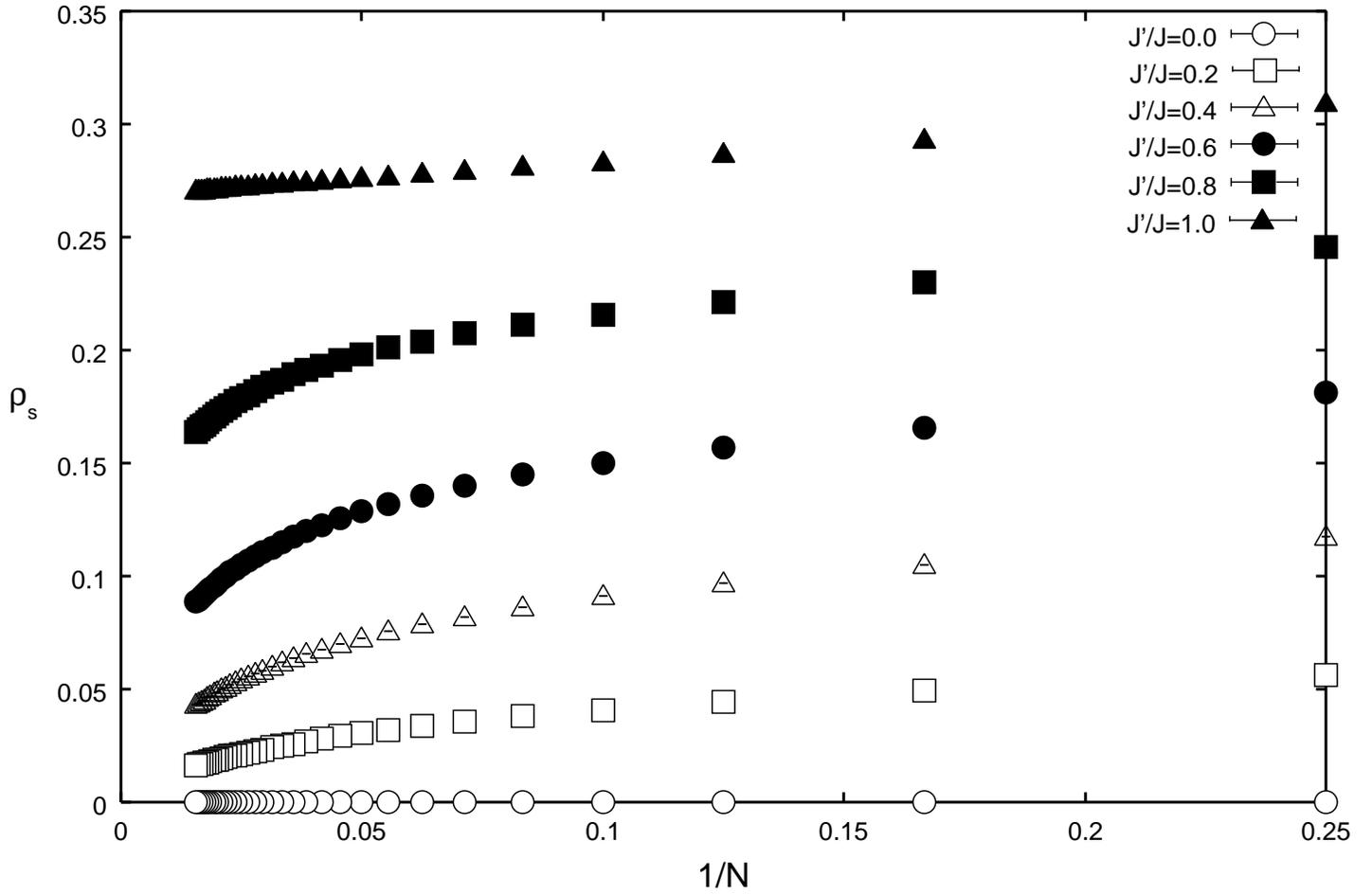}}
\caption{The stiffness factor $ \rho_s $ as a function of $ 1/N $, for lattice sizes $ N = 4 $ to 64.  }
\label{fig7}
\end{figure}

Putting together both the periodic and anti-periodic results for the energy, 
we can calculate the spin stiffness factor, given by 
\begin{equation}
\rho_{s} = \frac{2N}{\pi^{2}J}(E_0(N;\mbox{anti-periodic}) - E_0(N;\mbox{periodic}))
\end{equation}
The results are shown in Fig. \ref{fig7}. At couplings other than $J'/J = 1$ 
the values trend steadily down towards zero as the lattice size $ N $  
increases. There is a marked difference in behavior for the isotropic case 
$J'/J = 1$, as the strong curvature towards zero is not apparent. One would 
naively expect to obtain a bulk limit by a simple linear extrapolation 
procedure, but in fact the work of Woynarovich and Eckle \cite{woynarovich} has
 shown that there are logarithmic corrections to the ground state energy, and 
hence the stiffness. We can see the effects of these corrections for the 
$ J'/J = 1$ case, as there exists an exact result obtained by Hamer, Quispel and Batchelor \cite{hamer} (equation
(3.37) of ref. \cite{hamer} with $\gamma =0$)
\begin{equation}
\rho_{s} = \frac{1}{4} .
\label{eq3.2}
\end{equation}
This does not seem in accord with the data in Fig \ref{fig7}, which
appear to be approaching $0.27$. A late, logarithmic  downturn must therefore occur at very
large lattice sizes.

\begin{figure}
\centerline{\epsffile{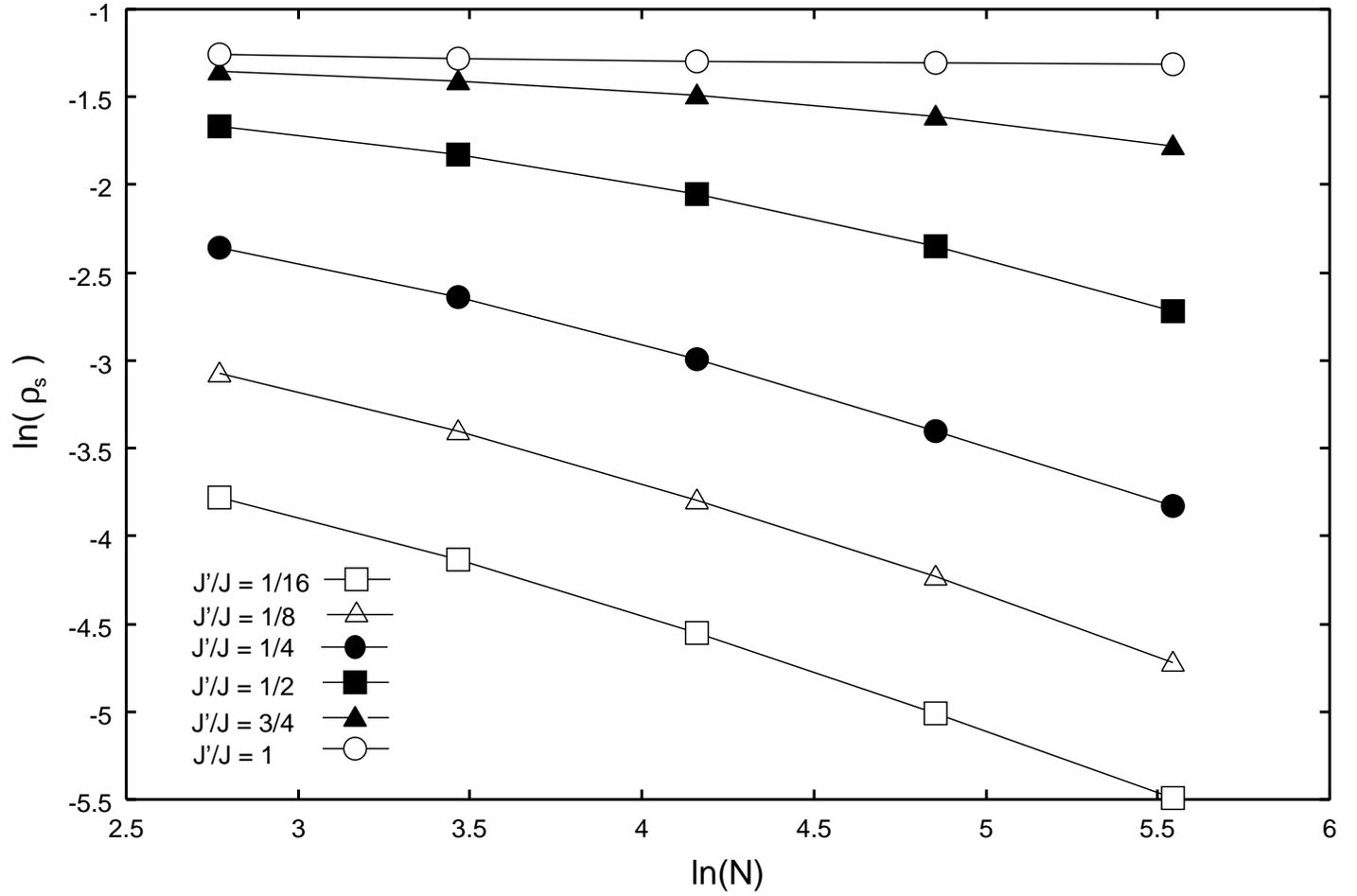}}
\caption{SSE results for the stiffness factor $ \rho_s $ versus lattice sizes ($N = 16,32,64,128,256 $). }
\label{fig8}
\end{figure}

\begin{figure}
\centerline{\epsffile{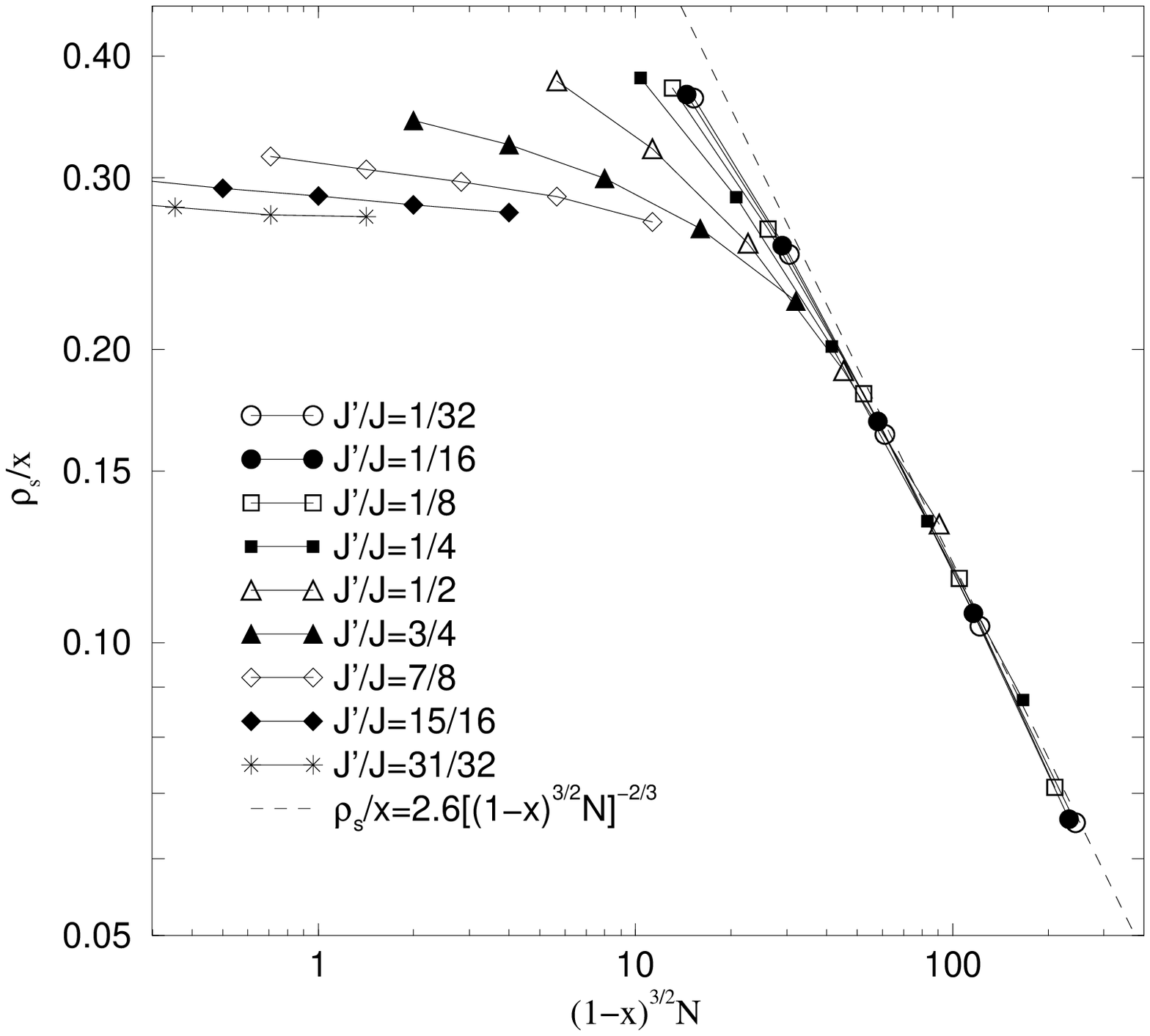}}
\caption{Scaling plot of $\rho_{s}/x$ versus $(1-x)^{3/2} N$, where $ x = J'/J $. Also shown is the scaling form (\ref{scaling}), which agrees with the data for large $ N $.}
\label{fig8a}
\end{figure}

Using field theoretical methods, Eggert and Affleck \cite{eggert} 
have predicted that a 
chain with $J' <  J$ should be similar to an open chain (i.e., $J'/J = 0$).
This implies that the spin [charge] stiffness should vanish as the system
size $N \to \infty$. Fig. \ref{fig8} shows SSE results\cite{footnote} for
the stiffness versus the system size ($N=16,32,64,128$ and $256$) for several
values of $J' \le J$. The results are in accordance with a scaling
behavior
\begin{equation}
\rho_{N}(x) \sim a(x)N^{-\sigma}
\end{equation}
where $x = J'/J$, and the index $\sigma \simeq 2/3$. Fig. \ref{fig8a}
demonstrates that the data for large $ N $ can be well described, in fact,
by a simple scaling form
\begin{equation}
\label{scaling}
\rho_{N}(x) \sim \frac{2.6x}{(1-x)}N^{-2/3}, \hspace{5mm} N \rightarrow
\infty.
\end{equation}
It is likely, however, that the true asymptotic correction-to-scaling
behavior is again being disguised by logarithmic corrections.

\begin{figure}
\centerline{\epsffile{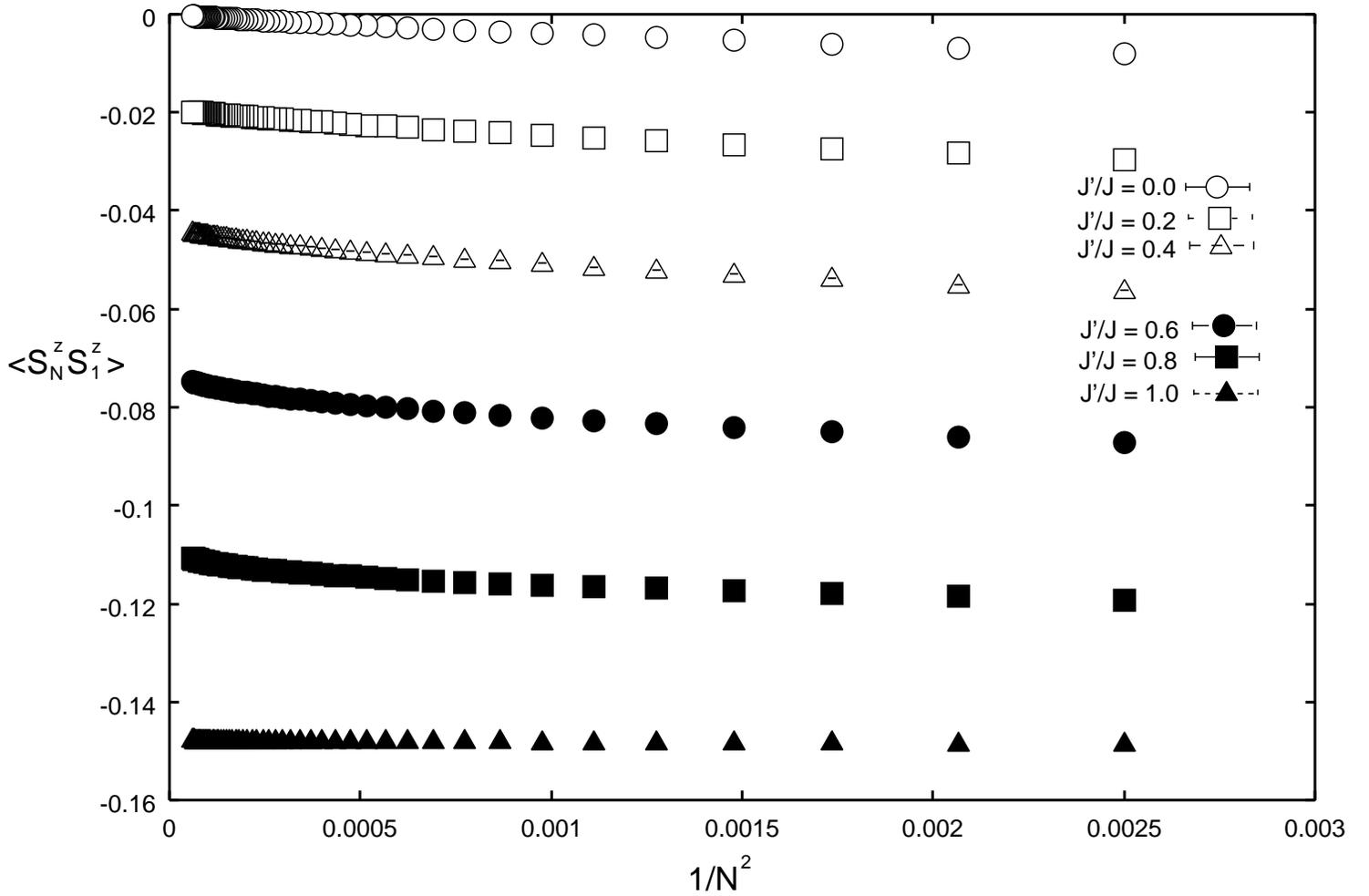}}
\caption{The correlation $ \langle S^z_N S^z_1 \rangle $ across the weak link as a function of $ 1/N^2 $. }
\label{fig9}
\end{figure}

We have also calculated the value of the spin correlation function
across the weak link, i.e. $ \langle S^z_N S^z_1 \rangle $. 
Fig. \ref{fig9} shows the behavior versus $1/N^2 $ for various
values of $J'/J$. 
It can be seen that the finite-lattice values
generally approach a finite value in the bulk limit, as one would
expect, except for the special case $J'/J = 0 $ where the link is open.
Theoretical expectations \cite{affleck} are that the correlation
function should approach its bulk limit like $1/N^{2}$, up to
logarithmic corrections. 

The presence of a finite correlation across the weak link is just what
one would naively expect when the weak-link coupling $J'$ is non-zero.
On the other hand, it might appear to contradict the previous statement
that a chain with a weak link should renormalize to the open chain. The
point here is that the weak-link correlation is a {\it local} quantity,
not a bulk property. It is only bulk properties such as the
spin-stiffness which scale to the value of the open chain.

\section{CONCLUSIONS} 

In summary, we have performed a finite-lattice study of the Heisenberg
ring with a weak link, using both DMRG and QMC calculations on rings of
up to 256 sites. The spin or `charge' stiffness has been calculated either
directly (QMC), or from
the energy difference between the system with anti-periodic boundary
conditions and that with periodic boundaries, assuming a quadratic
dependence of the energy on the twist parameter $\phi$ (DMRG). 

The stiffness, and hence the persistent current, is found to scale 
to zero in the bulk limit $N \rightarrow
\infty$, for any $J' < J$. This agrees with the renormalization group
prediction of Eggert and Affleck\cite{eggert}, 
that the stable fixed point for
this system corresponds to an open chain, so that the chain with a weak
link will behave like an open chain, as regards its bulk properties.

We have also measured the spin-spin correlation across the weak link. A
finite antiferromagnetic correlation remains in the bulk limit,
depending on the coupling $J'$ as one would expect. The renormalization
group argument does not apply to a `local' quantity such as this.

The asymptotic scaling behavior of these quantities has been disguised
to some extent by logarithmic finite-size scaling corrections. Eggert
and Affleck\cite{eggert} have circumvented this problem by studying a modified
model with an extra marginal operator; but we have not found this
necessary for our present purposes.

For the future, it would be of interest to see how the results
generalize to  more complicated and interesting cases, such as 
 higher spin chains,
or real electronic
models, such as the Hubbard model or its variant, the
so--called $t-J$ model.
Another interesting extension of the present study would be to interpret
the weak link and hence the modified bond in our model as caused by a
mechanical force on a quantum wire. It would be interesting to see what
conclusions could be drawn from our simple one-dimensional model for such a 
scenario.

\section{ACKNOWLEDGEMENTS}

We would like to thank Prof. I. Affleck for very useful correspondence.
This work was supported by a grant from the Australian Research Council,
and H.-P.E. was supported by an ARC-IREX Exchange Fellowship. A.W.S. was 
supported by the Academy of Finland (project 26175), and would also like 
to thank the School of Physics at UNSW for financial support through a 
Gordon Godfrey visitor fellowship. We are
grateful for the computing resources provided by the Australian Centre
for Advanced Computing and Communications (ac3) and the Australian
Partnership for Advanced Computing (APAC) National Facility.

\begin{table}
\caption{DMRG estimates of the
ground state energy $E_0/J$ and correlation $ \langle S^z_N S^z_1 \rangle $ for $ N = 64 $ sites at $ J'/J = 0.5 $ as a function 
of $m$, the
number of states retained per block. A SSE
estimate of $ E_0/J = -28.2178(3)$ for PBC agrees very well with the DMRG data: 
here our final DMRG estimate is -28.21797(1). The correlation between spins 
across the weak link also converges to better than 1 part in $10^4$. 
The anti-periodic boundary conditions yield similar levels of accuracy. }
\label{table1}
{\begin{tabular}{ccccc}
$ m $ & \multicolumn{2}{c}{Periodic} & \multicolumn{2}{c}{Anti-periodic} \\
      & $ E_0/J $ & $ \langle S^z_N S^z_1 \rangle $ & $ E_0/J $ & 
$ \langle S^z_N S^z_1 \rangle $ \\
\hline
96  & -28.217652 & -0.060888 &  -28.212779 & -0.0560186 \\
164 & -28.217938 & -0.061194 &  -28.213082 & -0.0562742 \\
234 & -28.217964 & -0.061219 &  -28.213123 & -0.0563028 \\
342 & -28.217970 & -0.061222 &  -28.213132 & -0.0563091 \\
\end{tabular}}
\end{table}
\begin{table}
\caption{
The internal energy and the spin stiffness calculated
in SSE simulations for a 256-site chain with $J'/J=1/4$ at different
inverse temperatures $\beta=J/T$.
}
\label{table2}
{\begin{tabular}{ccc}
$\beta$ & $E_0/NJ$ & $\rho_s$ \\
\hline
32   & -0.44211(2)  &  0.0000(0) \\
64   & -0.44237(1)  &  0.0008(1) \\
128  & -0.442429(9) &  0.0088(3) \\
256  & -0.442443(5) &  0.0202(5) \\
512  & -0.442443(4) &  0.0212(4) \\
1024 & -0.442453(3) &  0.0215(2) \\
2048 & -0.442448(3) &  0.0218(2) \\
\end{tabular}}
\end{table}

\center
\widetext
\input epsf


\begin{references}
\bibitem{footnote}
Note: the values obtained for the spin-stiffness by the DMRG and QMC
methods do not exactly agree, because the assumption of quadratic
behavior of the ground-state energy with $\Phi$ does not hold exactly
for a finite system. It does hold in the bulk limit, however, and the
results of the two methods converge in that limit.
\end{references}
\end{document}